\documentclass{article}

\usepackage{subfig}
\usepackage{booktabs, tabularx}

\usepackage[final]{neurips_robustseq_2022}

\usepackage[utf8]{inputenc} 
\usepackage[T1]{fontenc}    
\usepackage{url}            
\usepackage{booktabs}       
\usepackage{amsfonts}       
\usepackage{nicefrac}       
\usepackage{microtype}      
\usepackage{xcolor}         
\usepackage{graphicx}
\usepackage[colorlinks=true,linkcolor=blue!30!black,citecolor=blue!30!black,urlcolor=blue!30!black]{hyperref}

\title{Operationalizing Specifications, In Addition to Test Sets for Evaluating Constrained Generative Models}

\author{%
Vikas Raunak \quad Matt Post \quad Arul Menezes \\
Microsoft Azure AI\\
\texttt{\{viraunak,mpost,arulm\}@microsoft.com}\\}

\begin{document}

\maketitle

\begin{abstract}
In this work, we present some recommendations on the evaluation of state-of-the-art generative models for constrained generation tasks. The progress on generative models has been rapid in recent years. These large-scale models have had three impacts: firstly, the fluency of generation in both language and vision modalities has rendered common \textit{average-case} evaluation metrics much less useful in diagnosing system errors. Secondly, the same substrate models now form the basis of a number of applications, driven both by the utility of their representations as well as phenomena such as in-context learning, which raise the abstraction level of interacting with such models. Thirdly,  the user expectations around these models and their feted public releases have made the technical challenge of out of domain generalization much less excusable in practice. Subsequently, our evaluation methodologies haven't adapted to these changes. More concretely, while the associated utility and methods of interacting with generative models have expanded, a similar expansion has not been observed in their evaluation practices. In this paper, we argue that the scale of generative models could be exploited to raise the abstraction level at which evaluation itself is conducted and provide recommendations for the same. Our recommendations are based on leveraging specifications as a powerful instrument to evaluate generation quality and are readily applicable to a variety of tasks.
\end{abstract}

\section{Introduction}
\label{intro}

Recent advances in generative models across different modalities have enabled a myriad of applications, making the reliability failures of such models a frontier of research. Adjacent advances such as prompting \citep{gpt3}, open-vocabulary classification \citep{clip}, etc. have significantly enhanced the utility of such models, implicitly raising the abstraction level of human-model interactions. We believe that modern generative models have had three main impacts:
\begin{enumerate}
    \item \textbf{Fluent Generations}: The fluency of generation in both language and vision modalities has rendered existing benchmarks and metrics much less useful in diagnosing system errors \citep{gemv2, bigbench}, with previously widely used benchmarks and metrics on tasks such as machine translation \citep{salted} or summarization \citep{goyalzeroshotnews2022} becoming less useful in gauging system problems. 
    \item \textbf{Foundational Role}: The same substrate models now form the basis of a number of applications \citep{foundation_models} and this is driven both by the utility of their representations as well as phenomena such as in-context learning \citep{gpt3}, which raise the abstraction level of interacting with such models.
    \item \textbf{Greater User Expectations}: The user expectations around these models have made the technical challenge of out of domain generalization much less excusable in practice. 
\end{enumerate}

Subsequently, our evaluation methodologies haven’t adapted to these changes and the evaluation of modern generative models remains a challenge, with many hitherto standard benchmarks and metrics becoming less and less useful with increasing model capabilities. 

\section{Specifications for Evaluating Generative Models}


In this paper, we posit specifications as a powerful instrument for the evaluation of large-scale generative models. The evaluation analogy here is akin to the invention of fast transportation means, where platforms and guardrails had to be built for the society to safely and reliably use them. As these large-scale generative models enter more and more application domains, our primary concern here is about the observable behaviors of such models. And at this juncture, we think that the idea of specifications (as in engineering) is a more powerful way to think about the quality of these systems than the traditional evaluation practices in the machine learning community, primarily metrics and test cases. Besides empirical observations of generative models' failure modes, a strong basis for proposing a specifications-based evaluation framework stems from the fact that tails of the data distributions are notoriously hard to model and errors could arise as catastrophic failures on certain specific inputs despite the model's high average-case performance \citep{heavy_tail, belinkov2018synthetic}. By proposing a specifications-based evaluation framework, we also aim to naturally quantify model reliability as adherence to specifications.

To be precise, specifications are expressions of user intent on program behaviors \citep{gulwani2017program}. In traditional machine learning evaluation, references are a common way to express such specifications. And fundamentally, metrics are measurements based on those references, which are designed to mimic (correlate highly with) human judgements. Thereby, using references is one concrete instantiation of leveraging specifications. But, we can actually jump up an abstraction level and use arbitrary specifications for evaluation. We think this is important for modern generative models, owing to the insufficiency of metrics for evaluating system quality as well as due to the limitations of a strictly test-case based evaluation in eliciting system errors comprehensively.

\subsection{Insufficiency of Metrics}

Even though the metrics we use are getting better \citep{bert_score, comet, sellam-etal-2020-bleurt}, the assumptions behind metrics are problematic. These assumptions include: joint fluency and adequacy modeling through a single-dimension of system quality and no explicit sensitivity to salient errors at the instance-level. These assumptions behind system quality evaluation are not only detrimental to the downstream model user, but a lack of multidimensional view into model quality also makes the system developers unaware of different error types during development. We argue that this is much more problematic for large scale models where one can not iterate too fast.

\subsection{Insufficiency of Test Cases}
Methodologies such as CheckList \citep{checklist}, which build test cases for behavioral testing of models don't generalize to generative tasks. There are multiple reasons as to why a curated test-case (ground-truth) based evaluation cannot scale for a comprehensive evaluation of large-scale generative models, namely: (a) there are multiple equally plausible outputs corresponding to the same input, (b) errors in state-of-the-art generative systems are rare and vast amounts of data are required to elicit long-tail errors, (c) the errors or undesirable behaviors produced by the models are highly contextual, and static or limited-diversity test cases cannnot capture this, (d) error distributions across system iterations (whether in data, tokenization, model or learning) change, making a cache of test cases (however adversarial, e.g. as in \citep{anli}) obsolete in the long-run.

Further, with a view towards addressing the limitations of present evaluation protocols, we believe that a specifications based framework for evaluation should address four concerns:

\begin{enumerate}
    \item \textbf{Instance-Level Measurements}: The evaluation framework should provide targeted measurements of specification violations at the instance-level, unlike metrics, which typically work at the the level of corpora or sets.
    \item \textbf{Scalabilily}: The framework should be scalable to address error rarity and consume only input data, i.e. it should have no requirement for references.
    \item  \textbf{Invariance to (unstable) model error distributions}: The framework should not rely on a cache of test cases and be amenable to evaluate models across data or modeling iterations.
    \item \textbf{Trustworthy Measurements}: The framework should yield measurements of specification violations in a trustworthy manner, avoiding any Type I errors.
\end{enumerate}


In addition to the above desired characteristics, an implicit feature of specifications based evaluation is that it raises the abstraction level at which we conduct evaluations for such models, providing much more flexibility for the inclusion of arbitrary measurements (which could be catered even towards specific evaluation datasets' attributes). However, this gain in flexibility comes at the cost of operationalizing such specifications for evaluation, which could be non-trivial. In the next section, we demonstrate a case study on Machine Translation (MT) where we operationalized arbitrary specifications successfully to elicit and reliably measure a number of previously invisible errors. Further, in section \ref{reliability}, we posit some reliability failures of generative models as avenues ripe for similar specifications based evaluation. In section \ref{design}, we present a few sketches for operationalizing specifications on different generative model applications.





\begin{table*}[ht]
    \begin{tabularx}{\linewidth}{l X}
        \toprule
    \textbf{Property} & \textbf{Source-Translation Instance}  \\ \midrule
    
Physical Unit & Teacher's hallway song and dance reminds students to stay \colorbox{yellow}{6 feet} apart. \\
& Lehrer Flur Lied und Tanz erinnert die Schüler zu bleiben \colorbox{orange}{6 Meter} auseinander. \\ \hline

Currency & Floorpops Medina Self Adhesive Floor Tiles,  \colorbox{yellow}{£14} from Dunelm - buy now \\
 &  Floorpops Medina selbstklebende Bodenfliesen, \colorbox{orange}{15 €} von Dunelm günstig kaufen \\ 

\bottomrule
    \end{tabularx}
    \caption{Specification based evaluation for Google Machine Translation \citep{salted}}
    \label{tab:sota_examples}
    \vspace{-1.0em}
\end{table*}

\section{Operationalizing Specifications for Evaluation: A Case Study on MT}
\label{case_study}

In this section, we describe a case study on MT, where operationalizing a specification based framework for evaluation yielded measurements to quantify previously invisible errors in state-of-the-art systems. At a high-level, we start with specifications of correct behavior and then build \textit{detectors}, which check for violations of such specifications. The behavior specification is expressed for arbitrary input attributes or output properties. Some of these input attributes as such as physical units are at the token level, while some properties such as coverage are at the sequence level. More formally, the operationalization of an arbitrary specification as a trustworthy measurement is done is through a detector, which is an algorithm (iteratively constructed with human-in-the-loop) that, given an input-output instance, returns a boolean value indicating the presence of a specification violation with very high precision. We build detectors for 7 different specifications in MT \citep{salted}.

We find that by just passing a large number of input-output instances through these detectors, we gained quantifiable visibility into problems that were previously eliding measurements, such as translation of salient content or catastrophic errors such as hallucinations. For example, we find that errors such as incorrect conversion of physical units or currencies could be discovered and quantified at-scale using a specification which checks if the units/currencies have been carried over without any semantic change (an example is presented in Table \ref{tab:sota_examples}). Further, we find that even though such errors are rare, they are quite pervasive across state-of-the-art systems, posing a reliability threat for translation models. We believe that this approach to evaluation is well suited towards measuring reliability problems in other constrained generative models as well, which often fail in similar ways.

To summarize, our recommendation here is that by starting with an explicit role for specifications in evaluation, the effective evaluation set can be expanded dramatically by building error detectors which implement those specifications and generate trustworthy measurements. Such an evaluation effectively augments average-case performance metrics with specification based measurements, which could reliably quantify worst-case model behaviors during system development and beyond.

\section{Reliability Challenges of Large Scale Generative Models}
\label{reliability}

Throughout, this paper we posit \textit{reliability} in terms of \textit{adherence to certain specifications}. Under this view, we can interpret coarse-to-fine evaluation as going from looser specifications toward tighter ones for characterizing a model's failure modes. This way of characterizing model errors seems like a departure from standard practice, where we are used to thinking about quality in terms of average-case performance metrics such as BLEU, ROUGE, etc. \citep{bleu, lin-2004-rouge}. However, the framing of generation quality in terms of specifications helps us unify several phenomena under a single pedagogy. For example, the simplest specification for an generative system is that given a valid input, it should not output something absolutely irrelevant, and hallucinations in any modality could be characterized as samples on which a constrained generation model breaks this basic specification. 

Further, below, we enumerate some of the reasons as to why the nature of errors in state-of-the-art generative systems necessitates an evaluation protocol in which average-case performance measures are augmented with specifications-based measurements. Specifically, the reasons include:


\begin{enumerate}
    \item \textbf{Memorization}: High-capacity neural models are known to memorize the long-tail and templatic/repeated data. This poses a reliability risk inherent in any generative application of large neural models \citep{raunak-etal-finding-memo}, including privacy risks \citep{carlini}.
    \item \textbf{Real-world Data Distributions}: In general, it is very hard to map model behavior under out-of-distribution data settings. Further challenges, such as neural models' vulnerability to noise \citep{belinkov2018synthetic} or adversaries \citep{adversarial1, adversarial2} necessitate an evaluation protocol which could be scaled to arbitrary sets of inputs.
    \item \textbf{Biases and Spurious Correlations}: Besides spurious correlations that threaten model reliability \citep{bugs}, biases in the models could also impact their utility and exacerbate harms \citep{buolamwini2018gender, Bolukbasi}. 
    \item \textbf{Lack of Abstractions for Model Editing}: Large or \textit{gigantic} models with huge training costs run the risk of becoming software monoliths, due to a lack of abstractions in understanding, debugging and updating such models. This itself presents a risk towards their utility in sensitive domains where rapid interactivity vis-à-vis different stakeholders is required. Error visibility during model development becomes even more important in such cases.
\end{enumerate}

Coupling these challenges with the fact that these models have significantly expanded the scope of consumption of generative technologies \citep{clip, gpt3}, it is implied that their evaluation protocol needs to be more comprehensive than traditional smaller scale models. 

\begin{figure}%
    \centering
    \subfloat[\centering Plausible Output]{{\includegraphics[width=4cm]{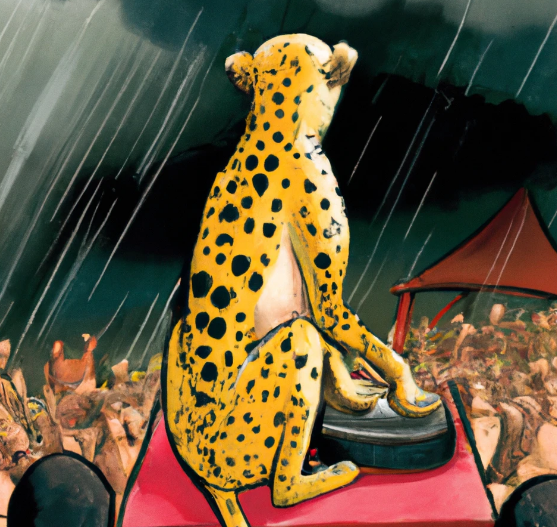} }}%
    \qquad
    \subfloat[\centering Implausible Output]{{\includegraphics[width=3.9cm]{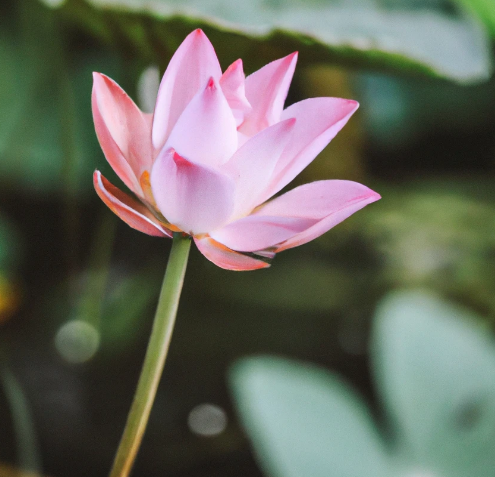} }}%
    \caption{An example of specification violation with DALL-E. The two images are generated from the input "\textit{cheetah is playing tabla on stage while the crowd is cheering on a rainy day}". Behaviors such as this could be enumerated by conducting reference-free specifications-based evaluation.}%
    \label{fig:example}%
    \vspace{-1em}
\end{figure}

\section{Sketches for Operationalizing Specifications}
\label{design}

In this section, we present a few sketches for applying specifications based evaluation for the evaluation of generative models, under some well known applications:
\begin{enumerate}
    \item \textbf{Text to Image Generation}: The design of specifications becomes considerably harder and compute-intensive, but still feasible when the transduction modalities change. Consider for example, Figure \ref{fig:example}, which shows two images generated by DALL-E \citep{dalle} for the same input text. Here, in case (b), the basic specification of the output having some support in the input is violated. For detecting such specification violations, an object detector could flag this instance as an error by checking for the object label in the text. And we can build trustworthy measurements by making this algorithm high precision.
    \item \textbf{Constrained Text Generation}: One of the harder cases to tackle for designing specifications is the case of constrained text generation for tasks which have no strong alignment between the input and output (unlike MT), such as data-to-text generation. However, even in such cases, many safety and reliability attributes are amenable to specifications based evaluation. Consider the safety violation involving gratuitous toxic text generation from the models; in this case, the evaluation set can be dramatically expanded through a specifications based approach which checks for toxic text attributes in generations from \textit{arbitrary} inputs.
\end{enumerate} 

\section{Leveraging Specifications for Human Evaluations}
\label{measurements}
Earlier, we argued that in order to obtain a trustworthy and multi-dimensional view into system quality during model evaluation or development, the procedure for checking specification violation (detector) be made high-precision. This allows generation of trustworthy error statistics at an arbitrary scale, allowing evaluation at scale without the need for any references. This simple high-precision rule of thumb for building measurements from specifications could then be leveraged for system evaluations, system comparisons or targeted evaluation of particular data or modeling interventions. For example, the specification of the output having some support in the input in the case of image generations automatically becomes a measurement of hallucinations, if its implementation, a hallucination detector (which checks for the violation of the specification in an instance) is made high-precision.



In this section, we propose some ideas on how new (arbitrary) categories of measurements designed for targeted evaluation of specific properties could be directly leveraged to guide human evaluations at different granularities. The idea of using fine-grained specifications to structure large-scale human evaluation has been explored in some of the more mature applications such as MT. For example, by grounding human evaluation in categories developed through error analysis for MT, the MQM framework in \citet{freitag2021experts} posits a hierarchy of translation errors which serve as annotation slots for human evaluation. Similarly, our key proposition for human evaluation is to leverage fine-grained specifications-based error analysis to explicitly guide the human annotation categories. Leveraging fine-grained specifications for human evaluation purposes could allow annotators to provide targeted signals which typically get lost in coarse grained categories. For example, the human evaluation in DALL-E was done on accuracy and realism axes. However, using categories grounded in specifications-based error analysis (or evaluation), new measurements such as 
omissions, additions, inconsistencies, etc. could be constructed to elicit and quantify more subtle effects in generation.

Further, even without the construction of an explicit error typology informed through the use of specifications, human evaluation can benefit by using specification violations as a sample selection step, i.e. the outputs flagged by using the detectors implemented to verify the specifications could be used by human evaluators as samples for system comparisons or targeted evaluations. This has the second-order effect of creating specialized test cases from inputs only, upon which further human evaluation could be conducted to elicit more characteristics of the models' failure modes.


\section{Summary and Conclusions}
\label{summary_and_conclusions}

To summarize, in this work, we presented some recommendations for the evaluation of generative models. We recommended evaluations based on explicitly formulated specifications, in order to raise the abstraction level of evaluation and allow it to scale arbitrarily. We believe such an evaluation framework could become increasingly important as large-scale constrained generative models enter real-world application domains, where safety and reliability problems are heavily penalized. To implement specifications, we proposed the use of high-precision as a rule of thumb for building trustworthy measurements. However, one limitation of this mode of operationalizing specifications is that not all errors could be readily enumerated, since the design itself necessitates a loss of recall. This could be partially mitigated by making the specifications very fine-grained. However, even specifications at coarser granularities (e.g., `coverage' in the MT case study \citep{salted}), will give visibility to error categories that are hard to elicit through curated test cases or through metric scores just through the advantage of \textit{evaluation scale}. Further, we presented how such specifications could further be leveraged to guide traditional human evaluations. Finally, we note that the design of specifications itself is a component of human evaluation, only at a higher abstraction level. 

\bibliographystyle{plainnat}
\bibliography{mybibfile}

\end{document}